\documentclass[letterpaper,conference]{IEEEtran}
\newcommand{\ms}{m_{\text{s}}}
\usepackage{amsfonts,  amssymb, color, textcomp}
\usepackage[all]{xy}
\usepackage[dvips]{graphicx}
\usepackage{latexsym}
\usepackage{epsfig}
\usepackage[cmex10]{amsmath} % Use the [cmex10] option
\begin{document}
\title{Free Distance Bounds for Protograph-Based Regular LDPC Convolutional Codes}

\author{
\authorblockN{David G. M. Mitchell$^*$, Ali E. Pusane$^\dag$, Norbert Goertz$^*$, and Daniel J. Costello, Jr.$^\dag$}
\authorblockA{$^*$Joint Research Institute
for Signal \& Image
Processing,
%School of Engineering and Electronics\\
The University of Edinburgh, Scotland,\\
\{David.Mitchell, Norbert.Goertz\}@ed.ac.uk\\
$^\dag$Dept. of Electrical Engineering, University of Notre Dame,
Notre Dame,
Indiana, USA,\\
\{apusane, dcostel$1$\}@nd.edu
}}

%\author{%
%\IEEEauthorblockN{David G. M. Mitchell$^*$, Norbert Goertz$^*$,\\ and Liam
%O'Carroll$^\dag$}
%\IEEEauthorblockA{$^*$Joint Research Institute for Signal and Image
%Processing,\\
%              $^\dag$Maxwell Institute for Mathematical Sciences,\\
%              The University of Edinburgh,\\
%              Scotland, UK,\\
%              \{David.Mitchell,Norbert.Goertz,L.O'Carroll\}@ed.ac.uk
%\and
%\IEEEauthorblockN{Ali E. Pusane, Kamil Sh. Zigangirov,\\ and Daniel J.
%Costello,
%Jr.}
%\IEEEauthorblockA{Dept. of Electrical Engineering,\\
%              University of Notre Dame,\\
%              Indiana, USA,\\
%              Email: \{apusane,kzigangi,dcostel1\}@nd.edu}
%}

\maketitle

\begin{abstract}
In this paper asymptotic methods are used to form lower bounds
on the free distance to constraint length ratio of several ensembles
of regular, asymptotically good, protograph-based LDPC convolutional
codes. In particular, we show that the free distance to constraint
length ratio of the regular LDPC convolutional codes exceeds that of
the minimum distance to block length ratio of the corresponding LDPC
block codes.
\end{abstract}

\section{Introduction}\label{sec:intro}
%Along with turbo codes, low-density parity-check (LDPC) block codes
%form a class of codes which approach the (theoretical) Shannon
%limit. LDPC codes were first introduced in the $1960$s by Gallager
%$\cite{gal}$. However, they were considered impractical at that time and very
%little related work was done until Tanner provided a graphical interpretation of
%the parity-check matrix in 1981 $\cite{tan}$. More recently, in his Ph.D.
%Thesis, Wiberg revived interest in LDPC codes and further developed the relation
%between
%Tanner graphs and iterative decoding \cite{wib}.

LDPC convolutional codes, the convolutional counterparts of LDPC
block codes, were introduced in \cite{fels}, and they have been
shown to have certain advantages compared to LDPC block codes of the
same complexity \cite{cost,cost2}. In this paper, we use ensembles
of $(J,K)$ regular tail-biting LDPC convolutional codes derived from
a protograph-based ensemble of LDPC block codes to obtain a lower
bound on the free distance to constraint length ratio of
unterminated, asymptotically good, periodically time-varying regular
LDPC convolutional code ensembles, i.e., ensembles that have the
property of free distance growing linearly with constraint length.

In the process, we show that the minimum distances of ensembles
of tail-biting LDPC convolutional codes (introduced in
\cite{zigvde}) approach the free distance of an associated
unterminated, periodically time-varying LDPC convolutional code
ensemble as the block length of the tail-biting ensemble increases.
We show that, for rate $1/2$ protograph-based ensembles with regular
degree distributions, the free distance bounds are consistent with
those recently derived for more general regular LDPC convolutional
code ensembles in \cite{srid} and \cite{truh}. Further, the
relatively low complexity requirements of computing the bound allows
us to calculate new free distance bounds that grow linearly with
constraint length for values of $J$ and $K$ that have not been
previously considered in the literature. We show, for all the
($J$,$K$)-regular ensembles considered, that the free distance to
constraint length ratio exceeds the minimum distance to block length
ratio of the corresponding block codes.

The paper is structured as follows. In Section \ref{sec:ldpccc}, we
briefly introduce LDPC convolutional codes. Section \ref{sec:proto}
summarizes the technique proposed by Divsalar to analyze the
asymptotic distance growth behavior of protograph-based LDPC block
codes \cite{div}. In Section \ref{sec:convfromproto}, we discuss
methods of forming regular convolutional codes from regular
protographs. We then describe the construction of tail-biting LDPC
convolutional codes as well as corresponding unterminated,
periodically time-varying LDPC convolutional codes in Section
\ref{sec:freedist}. In addition, we show that the free distance of a
periodically time-varying LDPC convolutional code is lower bounded
by the minimum distance of the block code formed by terminating it
as a tail-biting LDPC convolutional code. Finally, in Section
\ref{sec:bounds} we present lower bounds on the free distance of
ensembles of regular LDPC convolutional codes based on protographs.

\section{LDPC convolutional codes}\label{sec:ldpccc}
We start with a brief definition of a rate $R=b/c$ binary LDPC
convolutional code $\mathcal{C}$. (A more detailed description can
be found in \cite{fels}.) A code sequence $\mathbf{v}_{[0,\infty ]}$
satisfies the equation
\begin{equation}\label{codeword1}
\mathbf{v}_{[0,\infty ]}\mathbf{H}_{[0,\infty
]}^{\mathtt{T}}=\mathbf{0},
\end{equation}
where $\mathbf{H}_{[0,\infty]}^{\mathtt{T}}$ is the syndrome former matrix and

$\mathbf{H}_{[0,\infty]}=$ \vspace{1mm}\\
\hspace*{3mm}\scalebox{0.9}{\mbox{\scriptsize{$
\left[ \begin{array}{cccccc}
\mathbf{H}_{0}(0) & & & & \\
\mathbf{H}_{1}(1) & \mathbf{H}_{0}(1)& & & \\
\vdots & \vdots& & \ddots& \\
\mathbf{H}_{\ms}(\ms) & \mathbf{H}_{\ms-1}(\ms)&\ldots &\mathbf{H}_{0}(\ms) & \\
& \mathbf{H}_{\ms}(\ms+1) & \mathbf{H}_{\ms-1}(\ms+1)&\ldots
&\mathbf{H}_{0}(\ms+1) \\
&\ddots&\ddots&&\ddots
% \mathbf{H}_{0}^{\mathtt{T}}(t) &  & \cdots  &
%\mathbf{H}_{m_{s}}^{\mathtt{T}}(t+m_{s}) & \mathbf{0} &  &  \\
% \mathbf{0} & \mathbf{H}_{0}^{\mathtt{T}}(t+1) & \cdots  &  &
%\mathbf{H}_{m_{s}}^{\mathtt{T}}(t+m_{s}+1) & \ddots  &  \\
% \vdots  & \mathbf{0} & \ddots  & \vdots  & \vdots  & \ddots  &  \\
% &  & \ddots  & \mathbf{H}_{0}^{\mathtt{T}}(t') & \cdots  &  &
%\mathbf{H}_{m_{s}}^{\mathtt{T}}(t'+m_{s}) \\
\end{array} \right]$}
}}
\vspace{2mm}

\noindent is the parity-check matrix of the convolutional code
$\mathcal{C}$. The submatrices $\mathbf{H}_{i}(t)$,
$i=0,1,\cdots,\ms$, $t\geq 0$,  are binary $\left( c-b\right) \times c$
submatrices, given by
\begin{equation}
\mathbf{H}_i(t)= \left[ \begin{array}{ccc}
h_i^{(1,1)}(t) & \cdots & h_i^{(1,c)}(t) \\
\vdots & & \vdots \\
h_i^{(c-b,1)}(t) & \cdots & h_i^{(c-b,c)}(t) \\
\end{array} \right],\end{equation}
that satisfy the following properties:
\begin{enumerate}
\item $\mathbf{H}_{i}(t)=\mathbf{0},~i<0$ and $i>m_{s},~\forall ~t.$
\item There is a $t$ such that $\mathbf{H}_{\ms}(t) \neq \mathbf{0}.$
\item $\mathbf{H}_{0}(t)\neq\mathbf{0}$ and has full rank $\forall \,\, t$.
\end{enumerate}
We call $\ms$ the syndrome former memory and $\nu_{\text{s}} =
(\ms+1)\cdot c$ the decoding constraint length. These
parameters determine the width of the
nonzero diagonal region of $\mathbf{H}_{[0,\infty]}.$ %
The sparsity of the parity-check matrix is ensured by demanding that its rows
have very low Hamming weight, i.e., %
$w_{H}(\mathbf{h}_{i})<<(\ms+1)\cdot c,~i>0$, where $\mathbf{h}_{i}$ denotes the
$i$-th row of %
$\mathbf{H}_{[0,\infty ]}$. The code is said to be regular if its
parity-check matrix $\mathbf{H}_{[0,\infty]}$ has exactly $J$ ones
in every column and, starting from row $(c-b)\ms+1$, $K$ ones in
every row. The other entries are zeros. We refer to a code with
these properties as an $(m_\text{s},J,K)$-regular LDPC convolutional
code, and we note that, in general, the code is time-varying and has
rate $R=1-J/K$. A rate $R=b/c$, $(m_\text{s},J,K)$-regular
time-varying LDPC convolutional code is periodic with period $cT$ if
$\mathbf{H}_{i}(t)$ is periodic, i.e.,
$\mathbf{H}_{i}(t)=\mathbf{H}_{i}(t+T), \forall ~i,t$, and if
$\mathbf{H}_{i}(t)=\mathbf{H}_{i}, \forall ~i,t$, the code is
time-invariant.

%An LDPC convolutional code is called irregular if its row and column
%weights are not constant. The notion of degree distribution is used
%to characterize the variations of check and variable node degrees in
%the Tanner graph corresponding to an LDPC convolutional code.
%Optimized degree distributions have been used to design LDPC
%convolutional codes with good iterative decoding performance in the
%literature (see, e.g., \cite{zigvde,richter,arv,pus}), but no
%distance bounds for irregular LDPC convolutional code ensembles
%have been previously published.

\section{Protograph Weight Enumerators}\label{sec:proto}
\begin{figure}[htp]\begin{center}
\includegraphics[width=3.5in]{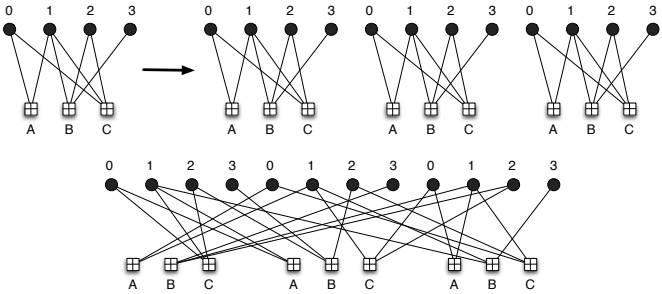}
\end{center}
\caption{The copy-and-permute operation for a
protograph.}\label{fig:proto}
\end{figure}

Suppose a given protograph has $n_v$ variable nodes and $n_c$ check nodes. An
ensemble of protograph-based LDPC block codes can be created using the
copy-and-permute
operation \cite{thor}. The Tanner graph obtained for one member of
an ensemble created with this method is illustrated in Fig.
\ref{fig:proto}.

%Note that the operation maintains node type connection, i.e., if a
%variable node from $\{0,1,2,3\}$ is connected to a check node from
%$\{A,B,C\}$ in the base protograph then the same connection must
%exist in the resulting protograph code.
The parity-check matrix $H$ corresponding to the ensemble of
protograph-based LDPC block codes can be obtained by replacing ones
with $N \times N$ permutation matrices and zeros with $N \times N$
all zero matrices in the underlying protograph parity-check matrix
$P$, where the permutation matrices are chosen randomly and
independently. The protograph parity-check matrix $P$ corresponding
to the protograph given in Figure \ref{fig:proto} can be written as
\begin{center}
\includegraphics[width=1.2in]{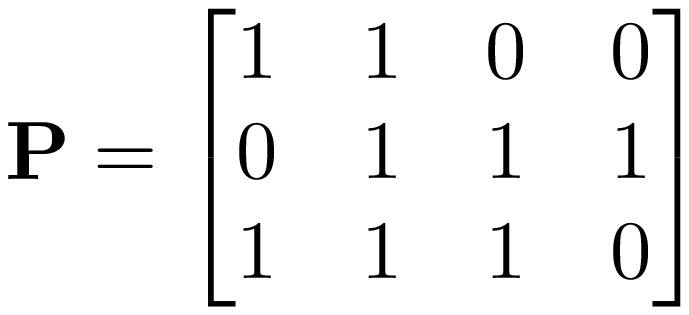}\raisebox{6mm}{,}
\end{center}
where we note that, since the row and column weights of $P$ are not
constant, $P$ represents the parity-check matrix of an irregular
LDPC code. For a $(J,K)$ regular LDPC code, the protograph contains $n_c=J$
check nodes and $n_v=K$ variable nodes, where each variable node is connected
to all $J$ check nodes, i.e., $P$ is an ``all-one'' matrix\footnote{It is also
possible to consider protograph parity-check matrices $P$ with larger integer
entries, which represent parallel edges in the the base protograph. In this
case, the corresponding block in $H$ consists of a sum of $N \times N$
permutation matrices. See \cite{thor} for details.}. 
%In the case when a
%variable node and a check node in the protograph are connected by $r$ parallel
%edges, the associated entry in $P$ equals $r$ and the corresponding block in
%$H$ consists of a summation of $r$ $N \times N$ permutation matrices. 
The sparsity condition of an LDPC parity-check matrix is thus satisfied for
large $N$. The code created by applying the copy-and-permute operation to
an $n_c\times n_v$ protograph parity-check matrix $P$ has block
length $n=Nn_v$. In addition, the code has the same rate and degree
distribution for each of its variable and check nodes as the
underlying protograph code.

Combinatorial methods of calculating ensemble average weight
enumerators have been presented in \cite{div} and \cite{fog}. We
now briefly describe the methods presented in \cite{div}.
\subsection{Ensemble weight enumerators}

Suppose a protograph contains $m$ variable nodes to be transmitted
over the channel and $n_v-m$ punctured variable nodes. Also, suppose that each
of the $m$ transmitted variable nodes has an associated weight $d_i \in
\{0,1\}$.
Let $S_d =\{(d_1,d_2,\ldots,d_m)\}$ be the set of all possible
weight distributions such that $d_1+\ldots+d_m=d$, and let $S_p$ be
the set of all possible weight distributions for the remaining punctured
nodes. The ensemble weight enumerator for the protograph is then given by
\begin{equation}
A_d = \sum_{\{d_k\}\in S_d}\sum_{\{d_j\}\in S_p}A_\mathbf{d},
\end{equation}
where $A_\mathbf{d}$ is the average number of codewords in the
ensemble with a particular weight distribution
$\textbf{d}=(d_1,d_2,\ldots,d_{n_v})$.

\subsection{Asymptotic weight enumerators}
The normalized logarithmic asymptotic weight distribution of a code
ensemble can be written as $r(\delta) = \lim_{n\rightarrow
\infty}\textrm{sup } r_n(\delta),$ where $r_n(\delta)
=\frac{\textrm{ln}(A_d)}{n}$, $\delta = d/n$, $d$ is the Hamming
weight, $n$ is the block length, and $A_d$ is the ensemble average
weight distribution.

%Normalize the weights $\epsilon_i=w_i/N$ and set $\sigma =
%(\epsilon_1+\epsilon_2+\epsilon_3)/2$. Then the asymptotic version
%of the number of sequences satisfying a check node is defined
%\cite{div} as
%\begin{equation}
%a^c(\epsilon_1,\epsilon_2,\epsilon_3) =
%H_3((\sigma-\epsilon_1),(\sigma-\epsilon_2),(\sigma-\epsilon_3)),
%\end{equation}
%where
%\begin{equation*}
%H_3(x_1,x_2,x_3) = -(1-\sum_{i=1}^3 x_i)\ln(1-\sum_{i=1}^3 x_i) -
%\sum_{i=1}^3 x_i \ln x_i,
%\end{equation*}
%and the normalized constraint
%$\max\{\epsilon_1,\epsilon_2,\epsilon_3\}\leq \sigma \leq 1$ is
%satisfied. The asymptotic weight enumerators for higher degree check
%nodes can be obtained by following this procedure for the
%concatenated non-asymptotic version.

% By using a change of variables, define
%$\tilde{r}_N(\tilde{\delta}) = \frac{\textrm{ln}(A_d)}{N}$ then as
%$N \rightarrow \infty$ we may obtain $r(\delta)$ as the asymptotic
%expression $\frac{1}{m}\tilde{r}(m\delta)$ where $m$ is the number
%of nodes to be transmitted. The following expression for
%$\tilde{r}(\tilde{\delta})$ was derived in \cite{div}:
%\begin{equation}
%\tilde{r}(\tilde{\delta}) = \max_{\{\delta_k\}\in
%S_{\bar{\delta}}}\max_{\{\delta_j\}\in S_{\pi}}\sum_{i=1}^{n_c}
%a^{c_i}(\mathbf{\delta}_i) - \sum_{i=1}^{n_v}(q_{v_i}-1)H(\delta_i),
%\end{equation}
%where $H(x)=-(1-x) \ln(1-x) - x \ln x$ is the entropy function. Note
%that the sets $S_{\bar{\delta}}$ and $S_{\pi}$ are normalized
%versions of $S_d$ and $S_p$ respectively (each component is divided
%by $N$ as $N \rightarrow \infty$).

Suppose the first zero crossing of $r(\delta)$ occurs at $\delta =
\delta_{min}$.  If $r(\delta)$ is negative in the range $
0<\delta<\delta_{min}$, then $\delta_{min}$ is called
the \emph{minimum distance growth rate} of the code ensemble. By
considering the probability
$$\mathbb{P}(d < \delta_{min} n) =
\sum^{\delta_{min}n-1}_{d=1}A_d,$$ it is clear that, as the
block length $n$ grows, if $\mathbb{P}(d < \delta_{min} n)
<<1$, then we can say with high probability that the majority of
codes in the ensemble have a minimum distance that grows linearly
with $n$ and that the distance growth rate is $\delta_{min}$.

\section{Forming Convolutional Codes from
Protographs}\label{sec:convfromproto} In this section, we present methods
to form convolutional parity-check matrices from the parity-check matrix of
a protograph.

\subsection{Unwrapping a protograph with gcd$(n_c,n_v)>1$}
Suppose that we have an $n_c\times n_v$ protograph parity-check
matrix $P$, where gcd$(n_c,n_v) = y>1$. We then partition $P$ as a
$y \times y$ block matrix as follows:
$$ P = \left[\begin{array}{ccc}
P_{1,1} &\ldots&P_{1,y}\\
\vdots&&\vdots\\
P_{y,1}&\ldots&P_{y,y}
\end{array}\right],$$
\noindent where each block $P_{i,j}$ is of size $n_c/y \times
n_v/y$. $P$ can now be separated into a lower triangular part,
$P_l$, and an upper triangular part minus the leading diagonal,
$P_u$. Explicitly,\vspace{1mm}
%\vspace{6mm}\hspace{-5mm} \scalebox{0.9}{\mbox{ \scriptsize{$P_l =
%\left[
%\begin{array}{cccc}
%P_{1,1} &&\\
%P_{2,1}&P_{2,2}&&\\
%\vdots&\vdots&\ddots&\\
%P_{y,1}&P_{y,2}&\ldots&P_{y,y}
%\end{array}
%\right] \textrm{ and } P_u = \left[\begin{array}{cccc}
%&P_{1,2}&\ldots&P_{1,y}\\
%&&\ddots&\vdots\\
%&&&P_{y-1,y}\\
%&&&
%\end{array}\right]$,}
%}}\vspace{2mm}

$$P_l =  \left[
\begin{array}{cccc}
P_{1,1} &&\\
P_{2,1}&P_{2,2}&&\\
\vdots&\vdots&\ddots&\\
P_{y,1}&P_{y,2}&\ldots&P_{y,y}
\end{array}
\right]$$and \vspace{2mm}$$P_u = \left[\begin{array}{cccc}
&P_{1,2}&\ldots&P_{1,y}\\
&&\ddots&\vdots\\
&&&P_{y-1,y}\\
&&&
\end{array}\right],$$
\noindent where blank spaces correspond to zeros. This operation is
called `cutting' a protograph parity-check matrix.

Rearranging the positions of these two triangular matrices and
repeating them indefinitely results in a parity-check matrix
$H_{cc}$ of an unterminated, periodically time-varying convolutional
code with decoding constraint length $\nu_\text{s}=n_v$ and period
$T=n_v$ given by
\begin{equation}\label{hcc}
H_{\textrm{cc}} = \left[\begin{array}{cccc}
P_l &&&\\
P_u&P_l&&\\
&P_u&P_l&\\
&&\ddots&\ddots
\end{array}\right].
\end{equation}
Note that the unwrapping procedure described above preserves the row and column
weights of the protograph parity-check matrix.
\vspace{1mm}

\subsection{Unwrapping a protograph with gcd$(n_c,n_v)=1$}\label{sec:gcd=1}
If gcd$(n_c,n_v) = 1$, we cannot form a square block matrix larger
than $1 \times 1$ with equal size blocks. In this case, $P_l = P$
and $P_u$ is the all zero matrix of size $n_c \times n_v$. This
trivial cut results in a convolutional code with syndrome former
memory zero, with repeating blocks of the original protograph on the
leading diagonal. We now propose two methods of dealing with this
structure.
\vspace{2mm}
\subsubsection{Form an $M$-cover}
Here, we create a larger protograph parity-check matrix by using the
copy and permute operation on $P$. This results in an $M n_c \times
M n_v = n_c^\prime \times n_v^\prime$ parity-check matrix for some
small integer $M$. The $n_c^\prime \times n_v^\prime$ protograph
parity-check matrix can then be cut following the procedure outlined
above. In effect, the $M \times M$ permutation matrix
creates a mini ensemble of block codes that can be be unwrapped to
an ensemble of convolutional codes. The resulting unterminated,
periodically time-varying convolutional code has decoding constraint
length $\nu_{\text{s}}=Mn_v$ and period $T=Mn_v$.
%\vspace{3mm}
\subsubsection{Use a nonuniform cut}
When gcd$(n_c,n_v)=1$, we can still form a convolutional code by
unwrapping the protograph parity-check matrix using a nonuniform
cut. Let the protograph parity-check matrix be written as
$$P=\left[\begin{array}{cccc}
p_{1,1} &\ldots&p_{1,n_v}\\
\vdots&&\vdots\\
p_{n_c,1}&\ldots&p_{n_c,n_v}
\end{array}\right].$$
We define a vector $\mathbf{\xi}$ consisting of $n_c$ step
parameters $\mathbf{\xi} = (\xi_1,\xi_2,\ldots,\xi_{n_c})$, where
$0\leq \xi_1<\xi_{n_c}\leq n_v$, and each $\xi_{i-1} < \xi_i$ for
$i=2,\ldots,n_c$
%\footnote{David, I need to have each $\xi_i$ to be
%strictly larger than $\xi_{i-1}$, $i=2,\ldots,n_c$ for the
%practicality paragraph to make sense. This does not seem to affect
%your examples which already satisfy this condition.}
. As in the
previous case, we form $n_c \times n_v$ matrices $P_l$ and $P_u$ as
follows
\begin{itemize}
 \item for each $\xi_i$, $i=1,\ldots,n_c$, the entries $p_{i,1}$ to
$p_{i,\xi_i}$ are copied into the equivalent positions in $P_l$;
\item entries $p_{i,\xi_i+1}$ to $p_{i,n_v}$ are copied, if they exist, into the
equivalent positions in $P_u$;
\item the remaining positions in $P_l$ and $P_u$ are set to zero.
\end{itemize}
We now form the parity-check matrix $H_{cc}$ of an unterminated,
periodically time-varying convolutional code as in (\ref{hcc}).
Nonuniform cuts do not change the row and column weights of the
parity-check matrix $P$. Further, the decoding constraint length
remains constant.

An LDPC convolutional code derived from an LDPC block code using a nonuniform
cut can be encoded and decoded using conventional encoding and decoding
methods with minor modifications. For an LDPC convolutional code obtained
using the nonuniform cut $\xi=(\xi_1,\ldots,\xi_{n_c})$, the maximum step width
$\xi_{max}$ for the cut is given by
\[
\xi_{max} = \max\limits_{i=2,\ldots,n_c}\{ \xi_1,\xi_i-\xi_{i-1}  \}.
\]
$\xi_{max}-\xi_i$ columns of zeros are then appended immediately to
the left of the columns in the original protograph parity-check
matrix $P$ corresponding to the steps $\xi_i$, $i=1,2,\ldots,n_c$, to
form a modified protograph parity-check matrix $P^\prime$. This
process is demonstrated for a (3,4)-regular protograph with the
nonuniform cut $\xi=(2,3,4)$ below:

\begin{center}
\includegraphics[width=\columnwidth]{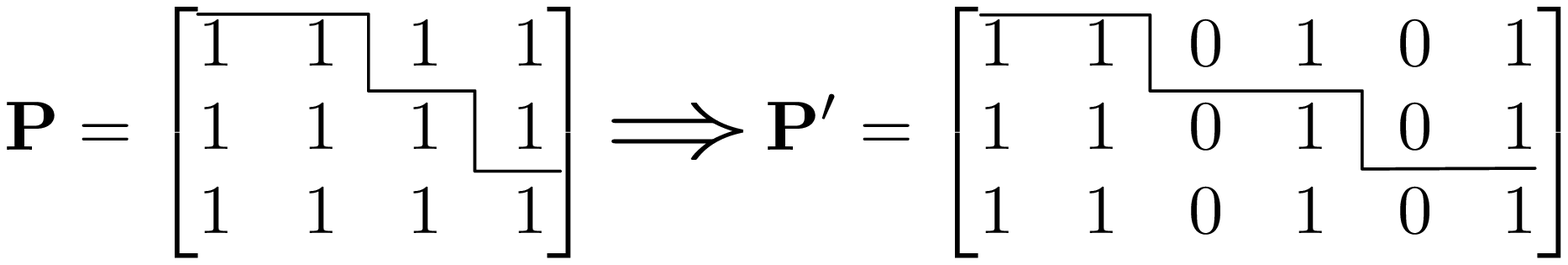}\raisebox{8mm}{.}
\end{center}

LDPC convolutional codes unwrapped from $P'$ can be encoded by a
conventional LDPC convolutional encoder with the condition that information
symbols are not assigned to the all-zero columns. Thus, these columns
correspond to punctured symbols, and the code rate is not affected. At the
decoder, a conventional pipeline decoder can be employed to decode
the received sequence. No special treatment is necessary for the
symbols corresponding to the all-zero columns, since the column
weight of zero insures that they are not included in any
parity-check equations, i.e., the belief-propagation decoding
algorithm ignores the corresponding symbols.
\section{Free distance bounds}\label{sec:freedist}
We now introduce the notion of tail-biting convolutional codes by
defining an `unwrapping factor' $\lambda$ as the number of times the
sliding convolutional structure is repeated. For $\lambda > 1$, the
parity-check matrix $H_{\textrm{tb}}^{(\lambda)}$ of the desired
tail-biting convolutional code can be written as

$$H_{\textrm{tb}}^{(\lambda)} = \left[\begin{array}{ccccc}
P_l &&&&P_u\\
P_u&P_l&&&\\
&P_u&P_l&&\\
&&\ddots&\ddots&\\
&&&P_u&P_l
\end{array}\right]_{\lambda n_c \times \lambda n_v}.$$

\noindent Note that the tail-biting convolutional code for $\lambda
= 1$ is simply the original block code.

\subsection{A tail-biting LDPC convolutional code ensemble}\label{sec:tailll}
Given a protograph parity-check matrix $P$, we generate a family of
tail-biting convolutional codes with increasing block lengths
$\lambda n_v$, $\lambda = 1,2,\ldots$, using the process described
above. Since tail-biting convolutional codes are themselves block
codes, we can treat the Tanner graph of
$H_{\textrm{tb}}^{(\lambda)}$ as a protograph for each value of
$\lambda$. Replacing the entries of this matrix with either $N
\times N$ permutation matrices or $N \times N$ all zero matrices, as
discussed in Section \ref{sec:proto}, creates an ensemble of LDPC
codes with block length $n=\lambda N n_v$ that can be analyzed
asymptotically as $N$ goes to infinity, where the sparsity condition
of an LDPC code is satisfied for large $N$. Each tail-biting LDPC
code ensemble, in turn, can be unwrapped and repeated indefinitely
to form an ensemble of unterminated, periodically time-varying LDPC
convolutional codes with decoding constraint length $\nu_s = N n_v$
and period $T=\lambda N n_v$.

Intuitively, as $\lambda$ increases, the tail-biting code becomes a
better representation of the associated unterminated convolutional
code, with $\lambda \rightarrow \infty$ corresponding to the
unterminated convolutional code itself. This is reflected in the
weight enumerators, and it is shown in Section \ref{sec:bounds} that
increasing $\lambda$ provides us with distance growth rates that
converge to a lower bound on the free distance growth rate of
the unterminated convolutional code.

\subsection{A free distance bound}
Tail-biting convolutional codes can be used to establish a lower
bound on the free distance of an associated unterminated,
periodically time-varying convolutional code by showing that the
free distance of the unterminated code is lower bounded by the
minimum distance of any of its tail-biting versions.
\newtheorem{freedist}{Theorem}
\begin{freedist}
Consider a rate $R=(n_v-n_c)/n_v$ unterminated, periodically
time-varying convolutional code with decoding constraint length
$\nu_\text{s}=Nn_v$ and period $T=\lambda N n_v$. Let $d_{min}$ be
the minimum distance of the associated tail-biting convolutional
code with length $n=\lambda N n_v$ and unwrapping factor
$\lambda>0$. Then the free distance $d_{free}$ of the unterminated
convolutional code is lower bounded by $d_{min}$ for any unwrapping
factor $\lambda$, i.e.,
\begin{equation}\label{emre}
    d_{free} \geq d_{min} ,\quad \forall \lambda > 0.
\end{equation}

\end{freedist}

\emph{Sketch of proof.} It can be shown that any codeword in a rate
$R=(n_v-n_c)/n_v$ unterminated, periodically time-varying
convolutional code with constraint length $\nu_\text{s}=N n_v$ and period
$T=\lambda Nn_v$ can be wrapped back to a codeword in a tail-biting
convolutional code of length $n=\lambda Nn_v$, for any $\lambda$. The
`wrapping back' procedure consists of dividing the convolutional
codeword into segments of length $\lambda N n_v$ and superimposing
these segments through a modulo summation. The Hamming weight of the
resulting codeword in the tail-biting code must be equal to or less
than that of the codeword in the unterminated code. The result is
then obtained by wrapping back the unterminated codeword with
minimum Hamming weight. For a formal proof, see \cite{truh}.

A trivial corollary of the above theorem is that the minimum
distance of a protograph-based LDPC block code is a lower bound on
the free distance of the associated unterminated, periodically
time-varying LDPC convolutional code. This can be observed by
setting $\lambda = 1$.

\subsection{The free distance growth rate}

The distance growth rate $\delta_{min}$ of a block code ensemble is
defined as its minimum distance to block length ratio. For the
protograph-based tail-biting LDPC convolutional codes defined in
Section \ref{sec:tailll}, this ratio is therefore given as
\begin{equation}
\delta_{min} = \frac{d_{min}}{n}=\frac{d_{min}}{\lambda N n_v}=\frac{d_{min}}{\lambda \nu_\text{s}}.
\end{equation}
Using (\ref{emre}) we obtain
\begin{equation} \label{zorro}
\delta_{min} \leq \frac{d_{free}}{\lambda \nu_\text{s}},
\end{equation}
where $d_{free}$ is the free distance of the associated
unterminated, periodically time-varying LDPC convolutional code. It
is important to note that, for convolutional codes, the length of
the shortest codeword is equal to the encoding constraint
length $\nu_\text{e}$, which in general differs from the decoding
constraint length $\nu_\text{s}$. Assuming minimal encoder and
syndrome former matrices, the relation between $\nu_\text{e}$ and
$\nu_\text{s}$ can be expressed as
\begin{equation} \label{spiderman}
\nu_\text{e} = \frac{1-R}{R} \nu_\text{s},
\end{equation}
which implies that, for code rates less than $1/2$, the encoding
constraint length is larger than the decoding constraint length.

Combining (\ref{zorro}) and (\ref{spiderman}) gives us the desired bound on the free distance growth rate
\begin{equation}
\delta_{free} \geq \frac{R}{1-R}\lambda \delta_{min},
\end{equation}
where $\delta_{free}=d_{free}/\nu_{e}$ is the convolutional code
free distance growth rate\footnote{If the syndrome former matrix is
not in minimal form, (\ref{spiderman}) results in an upper bound on
$\nu_\text{e}$, which implies that $\delta_{free}$ is underestimated
in this case.}.

\section{Bound computations}\label{sec:bounds}
We now present free distance growth rate results for ensembles of
asymptotically good, regular, LDPC convolutional codes based on
protographs. We consider all the regular ensembles originally
considered by Gallager \cite{gal}, and for each we calculate a lower
bound on the free distance to constraint length ratio
$\delta_{free}$. We begin by considering the regular ensembles with
gcd$(n_c,n_v)>1$. Then we consider the methods proposed in
Section \ref{sec:gcd=1} for regular ensembles with gcd$(n_c,n_v)=1$.
Results for these ensembles are then presented and discussed.
\subsection{Regular Ensembles with gcd$(n_c,n_v)>1$}
\begin{figure}[htp]\begin{center}
\includegraphics[width=3.3in]{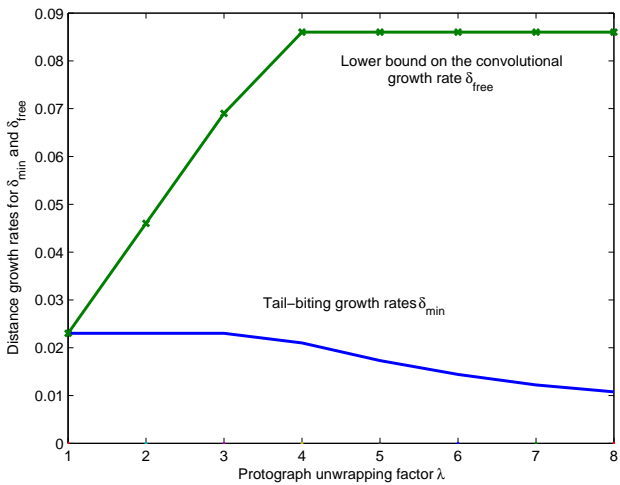}
\end{center}
\caption{Distance growth rates for Example $1$.}\label{fig:triv}
\end{figure}

\emph{Example $1$}: Consider the rate $R=1/2$, $(3,6)$-regular LDPC
code ensemble based on the following protograph:
\vspace{-1mm}
\begin{center}
\includegraphics[width=1.5in]{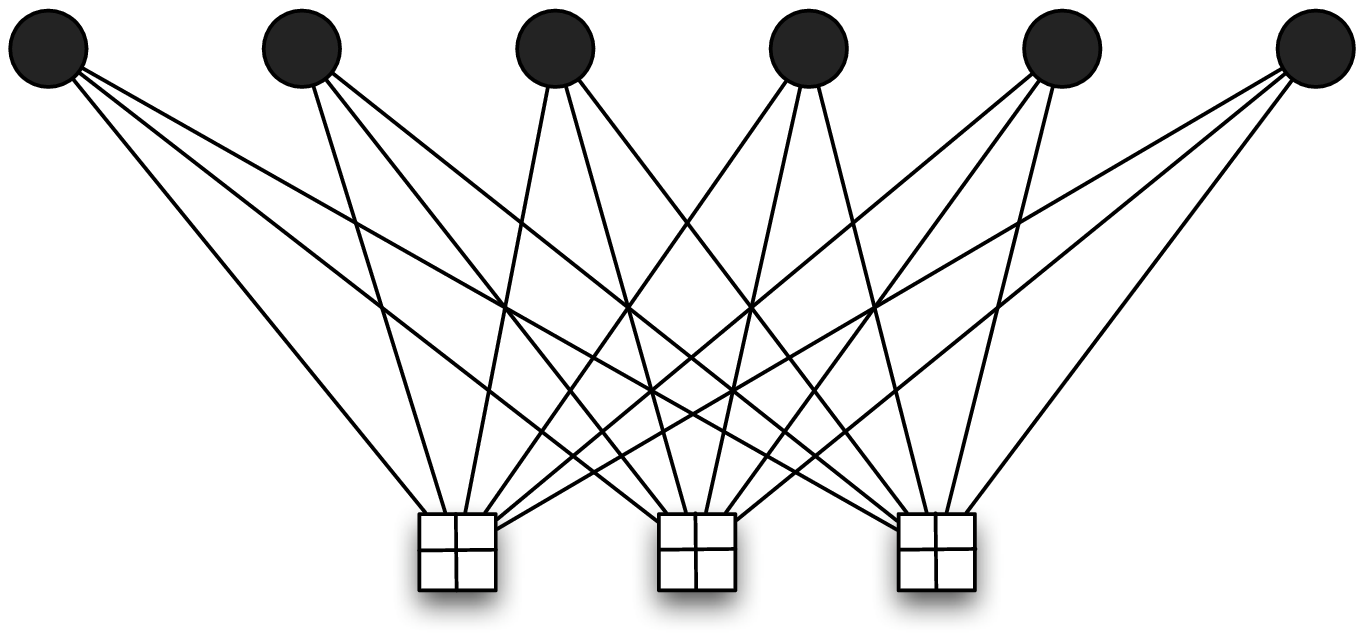}\raisebox{1cm}{.}
\end{center}
\vspace{-1mm}
\noindent For this example, the minimum distance growth rate is
$\delta_{min} = 0.023$, as originally computed by Gallager \cite{gal}.
%\begin{figure}[htp]\begin{center}
%\includegraphics[width=3.5in]{plot36.eps}
%\end{center}
%\caption{Asymptotic normalized weight distribution for $(3,6)$
%regular LDPC code.}\label{fig:36}
%\end{figure}
A family of tail-biting $(3,6)$-regular LDPC convolutional code ensembles can be
generated according to the following cut:
\vspace{-1mm}
\begin{center}
\includegraphics[width=1.5in]{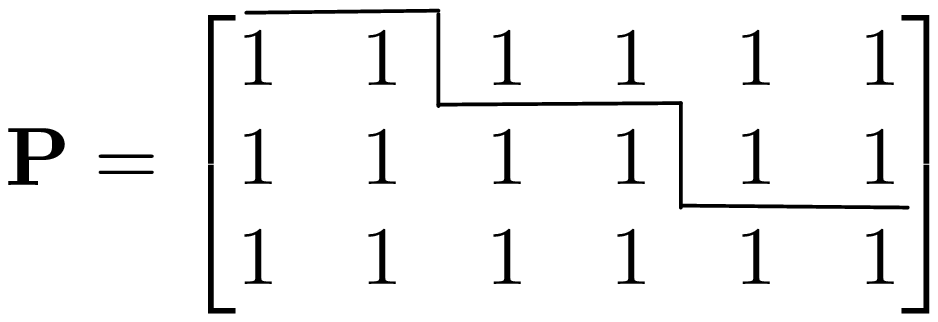}\raisebox{6mm}{.}
\end{center}
\vspace{-3mm}
For $\lambda=2,3,\ldots,8$, the minimum distance growth rate
$\delta_{min}$ was calculated for the tail-biting LDPC convolutional
code ensembles using the approach outlined in Section
\ref{sec:tailll}. These growth rates are shown in Fig.
\ref{fig:triv}, along with the corresponding lower bound on the free
distance growth rate $\delta_{free}$ of the associated ensemble of
unterminated, periodically time-varying LDPC convolutional codes.
For this rate $R=1/2$ ensemble, the lower bound on $\delta_{free}$
is simply $\delta_{free}\geq \frac{R}{1-R}\lambda
\delta_{min}=\lambda \delta_{min}$, since $\frac{R}{1-R}=1$ in this case.

We observe that, once the unwrapping factor $\lambda$ of the tail-biting
convolutional codes exceeds $3$, the lower bound on $\delta_{free}$ levels off
at $\delta_{free} \geq 0.086$, which agrees with the results presented in
\cite{srid} and \cite{truh} and represents a significant increase over the
value of $\delta_{min}$. In this case, the minimum weight
codeword in the unterminated convolutional code also appears as a
codeword in the tail-biting code\footnote{Example $1$ was
previously presented in \cite{isit08}.}.

\emph{Example $2$}: Consider the rate $R=1/3$, $(4,6)$-regular LDPC
code ensemble. The minimum distance growth rate for this ensemble is
$\delta_{min}=0.128$ \cite{gal}. We form a protograph in the usual fashion,
creating $4$ check nodes, each of which connect to all $6$ variable
nodes, and we observe that gcd$(4,6)=2$. The protograph parity-check
matrix and defined cut are displayed below:
\begin{center}
\includegraphics[width=1.5in]{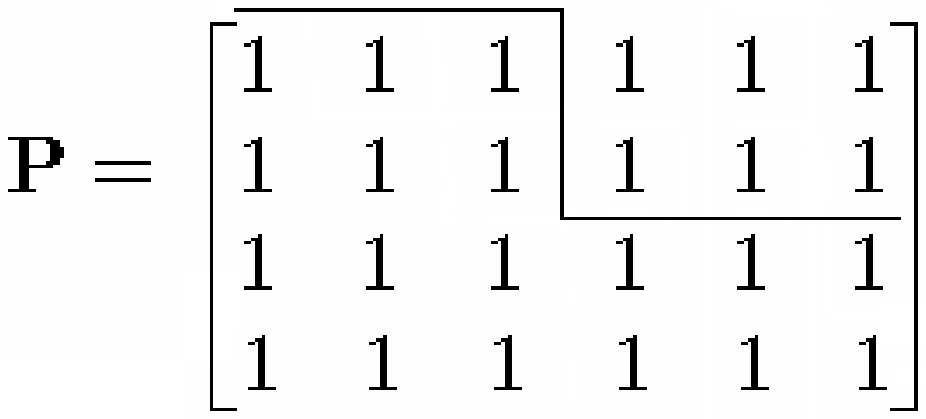}\raisebox{8mm}{.}
\end{center}

\noindent For this rate $R=1/3$ ensemble, the lower bound on
$\delta_{free}$ is $\delta_{free}\geq \frac{R}{1-R}\lambda
\delta_{min}=\frac{1}{2}\lambda \delta_{min}$. We observe that, as
in Example $1$, the minimum distance growth rates calculated for
increasing $\lambda$ provide us with a lower bound
$\delta_{free}\geq 0.197$ on the free distance growth rate of the
convolutional code ensemble, which exceeds the value of $\delta_{min}$.

\emph{Example $3$}: Consider the rate $R=1/2$, $(4,8)$-regular LDPC
code ensemble. The minimum distance growth rate for this ensemble is
$\delta_{min}=0.063$ \cite{gal}. The protograph parity-check matrix is cut
along the diagonal in steps of $1 \times 2$. For this rate $R=1/2$
ensemble, the lower bound on $\delta_{free}$ is $\delta_{free}\geq
\frac{R}{1-R}\lambda \delta_{min}=\lambda \delta_{min}$, and we
obtain the lower bound $\delta_{free}\geq 0.191$ on the free
distance growth rate of the convolutional code ensemble, which is again
significantly larger than $\delta_{min}$.
\subsection{Regular Ensembles with gcd$(n_c,n_v)=1$}
We now present results for the two methods of unwrapping a
protograph with gcd$(n_c,n_v)=1$ introduced in Section
\ref{sec:gcd=1}.

\begin{figure}[htp]
\begin{center}
\includegraphics[width=3.3in]{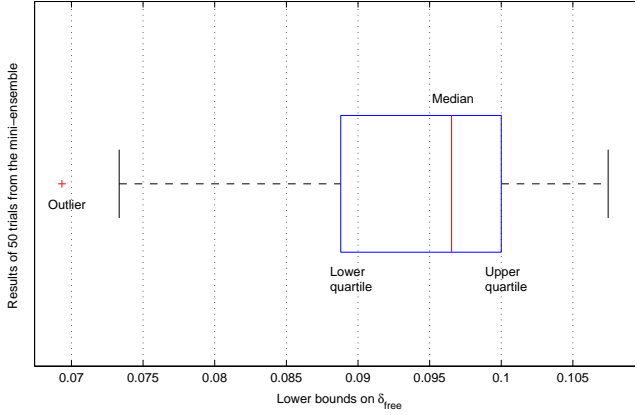}
\end{center}
\caption{Free distance growth rates for $50$ mini-ensemble
members.}
\label{fig:box}
\end{figure}

\emph{Example $4$}: Consider the rate $R=2/5$, $(3,5)$-regular
ensemble. The minimum distance growth rate for this ensemble is
$\delta_{min}=0.045$ \cite{gal}. For this rate $R=2/5$ ensemble, the lower
bound on $\delta_{free}$ is $\delta_{free}\geq \frac{R}{1-R}\lambda
\delta_{min}=\frac{2}{3}\lambda \delta_{min}$. The first approach
was to form a two-cover of the regular protograph. The resulting
mini-ensemble has $2^{n_vn_c}=2^{15}$ members. Fifty distinct
members were chosen randomly. The resulting lower
bounds calculated for $\delta_{free}$ are shown in a box plot in Fig.
\ref{fig:box}.

We observe a fairly large spread in results from the
mini-ensemble. The median from the fifty trials is
$\delta_{free}=0.097$. We also observe that the smallest lower bound found is
statistically a outlier as it lies reasonably far away from the
lower quartile. Note that this smallest lower bound ($\delta_{free}\geq
0.069$) is larger than the block code growth
rate $\delta_{min}=0.045$. Also, the best lower bound,
$\delta_{free}\geq 0.108$, is significantly larger than
$\delta_{min}$.

We now consider the nonuniform cut case. Consider the following two
nonuniform cuts of the standard protograph parity-check matrix for
the regular $(3,5)$ ensemble:

\vspace{-4mm}
\begin{figure}[htp]
\begin{center}
\includegraphics[width=1.3in]{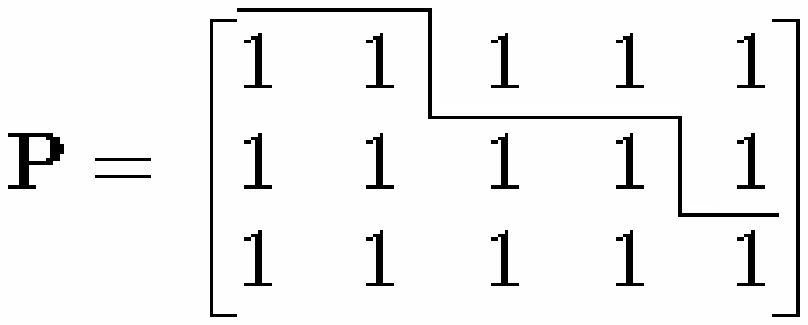}\raisebox{6.2mm}{\textrm{
and
}}\includegraphics[width=1.3in]{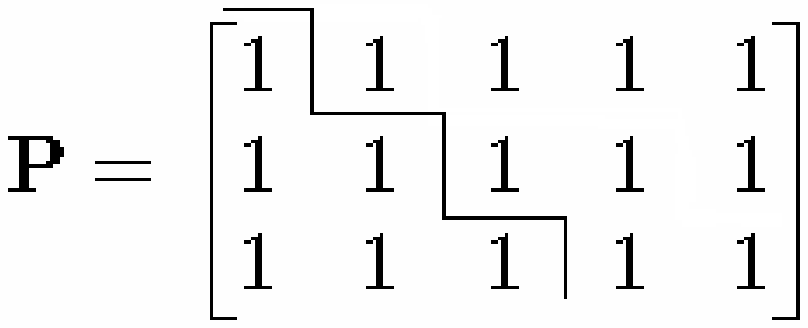}\raisebox{6.2mm}{\textrm{,}}
\end{center}
\end{figure}
\vspace*{-5mm}

 \noindent with corresponding cutting vectors
$\xi_1=(2,4,5)$ and $\xi_2=(1,2,3)$. We calculate a lower bound of
$\delta_{free}\geq 0.119$ for cut one and $\delta_{free}\geq 0.111$
for cut two. Both nonuniform cuts give larger lower bounds on
$\delta_{free}$ than the mini-ensemble method.

For the remaining regular ensembles with gcd$(n_c,n_v)=1$, we used
the nonuniform cut method. The resulting bounds are given in the
table below. \vspace*{2mm}

\noindent \scalebox{0.93}{\begin{tabular}{c|c|c|c} Ensemble & Cut $\mathbf{\xi}$
&
$\delta_{min}$ \cite{gal} & Lower bound on
$\delta_{free}$\\
\hline
$(3,4)$ & $(2,3,4)$ & $0.112$ & $0.177$\\
$(4,5)$ & $(2,3,4,5)$ & $0.210$ & $0.266$\\
$(5,6)$ & $(2,3,4,5,6)$ & $0.254$ & $0.317$
 \end{tabular}}
\vspace*{2mm}

\begin{figure}[htp]
\begin{center}
\includegraphics[width=\columnwidth]{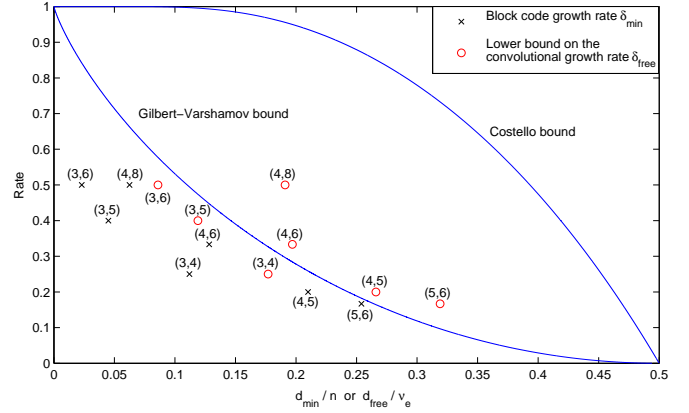}
\end{center}
\caption{Comparison of calculated growth rates with the
Gilbert-Varshamov bound for block code minimum distance growth rates and the
Costello bound for convolutional code free distance growth
rates.}\label{fig:gvc}
\end{figure}

\noindent For each ensemble considered, the lower bound on
$\delta_{free}$ is significantly larger than $\delta_{min}$ for the block code
ensemble. This is illustrated in Fig. \ref{fig:gvc}, where the distance
growth rates of each regular LDPC code ensemble are compared to the
corresponding bounds for general block and convolutional codes. \vspace{-2mm}

%\vspace{-0.5mm}
\section{Conclusions}
%\vspace{-1mm}
In this paper, asymptotic methods were used to
calculate a lower bound on free distance that grows linearly
with constraint length for several ensembles of regular,
unterminated, protograph-based periodically time-varying LDPC
convolutional codes. It was shown that the free distance growth
rates of the regular LDPC convolutional code ensembles exceed the
minimum distance growth rates of the corresponding regular LDPC
block code ensembles. When gcd$(n_c,n_v)=1$, we proposed two
new methods of unwrapping the protograph parity-check matrix in
order to obtain the best possible lower bound on $\delta_{free}$.
The results suggest that we typically obtain better lower bounds by
performing nonuniform cuts. %\vspace{-0.8ex}
\section*{Acknowledgement}
%\vspace{-0.8ex}
This work was partially supported by NSF Grants
CCR02-05310 and CCF05-15012 and NASA Grant NNX07AK536. In addition,
the authors acknowledge the support of the Scottish Funding Council
for the Joint Research Institute with the Heriot-Watt University,
which is a part of the Edinburgh Research Partnership.

\end{document}